\documentclass[footinbib,aps,pra,reprint,superscriptaddress]{revtex4-1}

\usepackage{amsmath,amssymb,bbm,bm,graphicx}
\usepackage{dirtytalk}
\usepackage{mathtools}
\usepackage{braket}
\usepackage[absolute]{textpos}

\usepackage{xcolor}

\def\cF{\mathcal{F}}
\def\cP{\mathcal{P}}
\DeclareMathOperator{\Tr}{Tr}

\begin{document}

\title{Bell state analyzer for spectrally distinct photons}

\author{Navin B. Lingaraju}
\email{navin.lingaraju@sri.com}
\thanks{These authors contributed equally to this work.}
\author{Hsuan-Hao Lu}
\thanks{These authors contributed equally to this work.}
\author{Daniel E. Leaird} 
\affiliation{School of Electrical and Computer Engineering and Purdue Quantum Science and Engineering Institute, Purdue University, West Lafayette, Indiana 47907, USA}
\author{Steven Estrella}
\affiliation{Freedom Photonics, LLC, 41 Aero Camino, Santa Barbara, CA 93117, USA}
\author{Joseph M. Lukens}
\affiliation{Quantum Information Science Group, Oak Ridge National Laboratory, Oak Ridge, TN 37831, USA}
\author{Andrew M. Weiner}
\affiliation{School of Electrical and Computer Engineering and Purdue Quantum Science and Engineering Institute, Purdue University, West Lafayette, Indiana 47907, USA}

%\ociscodes{(270.5585) Quantum information anFinad processing; (270.0270) Quantum optics; (320.5540) Pulse shaping; (060.5060) Phase modulation.}

\begin{abstract}
We demonstrate a Bell state analyzer that operates directly on frequency mismatch. Based on electro-optic modulators and Fourier-transform pulse shapers, our quantum frequency processor design implements interleaved Hadamard gates in discrete frequency modes. Experimental tests on entangled-photon inputs reveal accuracies of $\sim$98\% for discriminating between the $\ket{\Psi^+}$ and $\ket{\Psi^-}$ frequency-bin Bell states. Our approach resolves the tension between wavelength-multiplexed state transport and high-fidelity Bell state measurements, which typically require spectral indistinguishability.
\end{abstract}

\maketitle

\begin{textblock}{13.6}(1.25,15.1)\noindent\fontsize{6}{6}\selectfont\textcolor{black!30}{This manuscript has been co-authored by UT-Battelle, LLC, under contract DE-AC05-00OR22725 with the US Department of Energy (DOE). The US government retains and the publisher, by accepting the article for publication, acknowledges that the US government retains a nonexclusive, paid-up, irrevocable, worldwide license to publish or reproduce the published form of this manuscript, or allow others to do so, for US government purposes. DOE will provide public access to these results of federally sponsored research in accordance with the DOE Public Access Plan (http://energy.gov/downloads/doe-public-access-plan).}
\end{textblock}

%We demonstrate a Bell state analyzer capable of accommodating frequency mismatch – an enabling advance for mediating entanglement between heterogeneous nodes and leveraging the bandwidth optical fiber for spectrally multiplexed quantum state transport.

%Our approach accommodates joint measurements between spectrally distinguishable photons without any tradeoff between entanglement fidelity and the entanglement generation rate. Furthermore, this approach is based on a paradigm that supports parallel quantum operations that can be reconfigured with simple electronic control. This ability to accommodate frequency mismatch represents an enabling advance for mediating entanglement between heterogeneous nodes and leveraging the bandwidth optical fiber for spectrally multiplexed quantum state transport.
%We demonstrate the first Bell-state analyzer for frequency-bin entangled photons and unambiguously distinguish two of four Bell states with fidelities over 92\%. This is an important step toward the vision of a quantum internet that is compatible with both heterogeneous nodes and dense spectral multiplexing.

Unlocking the potential of quantum technology will require not just progress in developing standalone systems, but also in mediating communications and entanglement between these systems across a network~\cite{Wehner2018}. Protocols have been devised to herald the generation of remote entanglement between previously unentangled parties, often through Bell state measurements~\cite{Duan2003, Zhao2007, Sangouard2011}.  In the simplest conception of such a protocol, two photons, each entangled with separate qubits (either matter-based or photonic), are mixed at a 50:50 spatial beamsplitter. Quantum interference between two-photon outcomes leads to one of many detection events, a subset of which %, usually signified by two coincident detector clicks, 
heralds projection of the undetected qubits onto a \textit{known} entangled state. Similarly, in quantum teleportation~\cite{Bennett1993,Bouwmeester1997, Pirandola2015}, a Bell state measurement performed on an unknown input qubit and one half of an entangled pair projects the remaining qubit onto the state of the original input---up to a known unitary rotation---all without physical transmission of the quantum state itself.

However, a limitation of conventional Bell state measurements is that photons participating in the joint measurement need to be spectrally indistinguishable in order to project the remaining undetected particles onto a \textit{known} entangled state. For the case when photons are separated by a frequency difference $\Delta\omega$ and detected a time $\Delta t$ apart, the undetected qubits are projected onto an entangled state with an additional phase shift $e^{\mathrm{i}\Delta\omega \Delta t}$~\cite{Zhao2014,Vittorini2014}. While the frequency difference remains fixed, the term $\Delta t$ changes from one event to the next. Irrespective of whether one applies temporal postselection or active feedforward to compensate for this fluctuating term, the fidelity of remote entanglement is ultimately limited by the timing resolution of photon detection ($\Delta t_{R}$), which places a lower bound on phase accuracy on the order of $\Delta\omega \Delta t_{R}$. This presents a challenge to networks with heterogeneous nodes or those that rely on spectral multiplexing, as a $\pi$ phase uncertainty at a frequency difference of just $10$ GHz corresponds to $\Delta t_R = 50$~ps---placing a strong demand on detection jitter.

In this work, we demonstrate the first Bell state analyzer (BSA) that operates on frequency mismatch directly using electro-optic mixing techniques. Through the use of interleaved frequency beamsplitters in a quantum frequency processor~\cite{LuHOM}, we realize unambiguous measurement of two frequency-bin–encoded Bell states with a discrimination accuracy of 98\%. Our demonstration makes it possible to break the tradeoff between spectral distinguishability and remote entanglement fidelity, representing an important step toward the long-term vision of a quantum internet that is compatible with both heterogeneous nodes and dense spectral multiplexing.

\begin{figure*}[tb!]
\centering
  \includegraphics[width=6.4in]{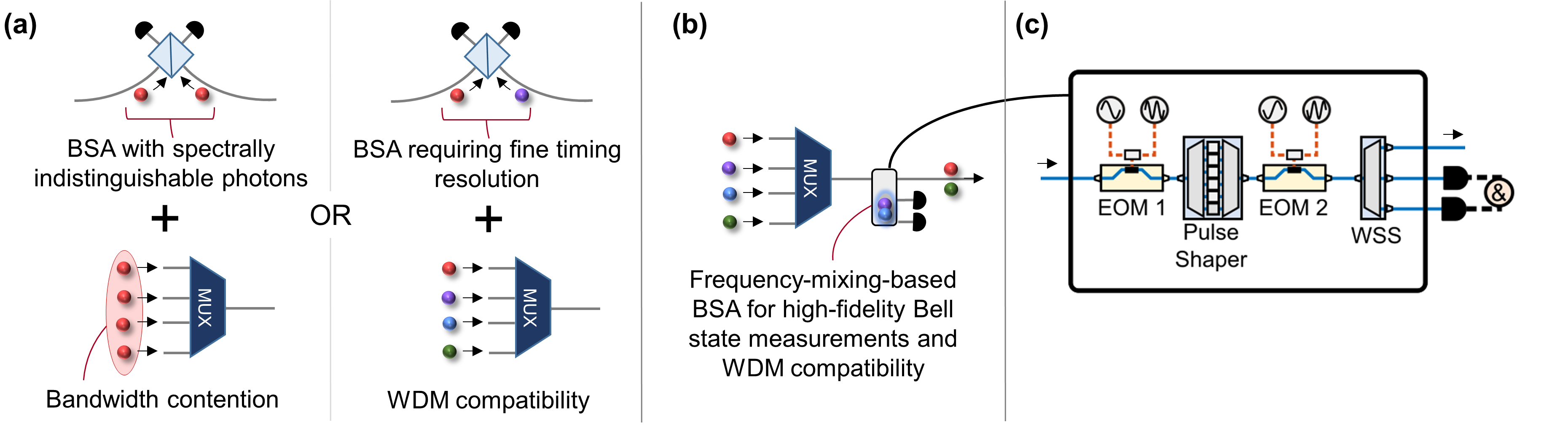}
  \caption{(a)~Conventional BSAs are incompatible with wavelength-division multiplexed networks, which support wavelength-selective aggregation and routing with multiplexing hardware (MUX). (b)~Frequency-mixing-based solution, in which all photons carry distinct frequencies, and Bell state measurements erase frequency mismatch directly. (c)~Configuration proposed and analyzed, %experimentally implemented, 
  based on a three-element QFP driven with dual-tone electro-optic modulation.}
  \label{fig1}
\end{figure*}

Figure~\ref{fig1}(a) highlights the challenges faced in implementing conventional spatial BSAs in wavelength-multiplexed fiber-optic networks. On the one hand, photons with identical carrier frequencies are optimal for high-fidelity Bell state measurements on a beamsplitter (or fiber-optic coupler), yet the selective adding and dropping of separate photons with identical spectra %, but adding and dropping spectrally identical photons in the network 
is incompatible with wavelength multiplexers, creating bandwidth contention issues. On the other hand, spectrally distinct photons can be readily multiplexed but require fine temporal resolution in the BSA to mitigate the distinguishability otherwise present with frequency mismatch, increasingly difficult for the tens of GHz spacings typical in fiber networks. Our proposed solution is outlined in Fig.~\ref{fig1}(b). Here, all photons enter the network on a distinct available wavelength channel; to implement a BSA, two photons are spectrally isolated and measured directly with a ``frequency-mismatch-erasing'' operation, depicted schematically as a gray box. In principle such an operation can be realized by nonlinear-optical mixing with classical pump fields~\cite{Raymer2010, Kobayashi2016, Clemmen2016, Joshi2020b}. In our case we leverage the flexibility of a quantum frequency processor (QFP), which is capable of synthesizing arbitrary unitary transformations in discrete frequency bins~\cite{Lukens2017, Lu2019c}. Experimentally, we focus on a three-element QFP as pictured in Fig.~\ref{fig1}(c), consisting of a pulse shaper sandwiched between two electro-optic phase modulators (EOMs). The EOMs are driven by a superposition of two microwave tones, equal to the fundamental frequency-bin spacing and its second harmonic. Such a two-tone setup matches the most expressive QFP hiterto demonstrated (in terms of the number of parameters available for design) which was previously explored to realize a frequency-bin tritter~\cite{Lu2018}---a $3\times3$ extension of the beamsplitter.

In general, the importance of eliminating frequency mismatch applies to photons carrying quantum information in any degree of freedom, such as polarization, path, or time bins. However, in light of the synergies of frequency encoding with both matter-based qubits that leverage multiple energy levels and fiber-optic networks built on wavelength multiplexing,
%However, for clearness of presentation and to best align with fiber-optic networks based on frequency bins, 
we design our proof-of-principle BSA for frequency-bin qubits specifically. Suppose that the photons to be measured exist in a superposition of frequency bins $\{A_0,A_1\}$ (qubit A) and $\{B_0,B_1\}$ (qubit B) selected from a predefined grid spaced in multiples of $\Delta\omega$. Following the conventional BSA for polarization qubits~\cite{Braunstein1995, Mattle1996}, the desired transformation should mix the logical-0 modes of both photons ($A_0$ and $B_0$) according to a 50:50 beamsplitter, i.e., Hadamard operation; the logical-1 modes ($A_1$ and $B_1$) should also be mixed according to an independent Hadamard gate. Two of the four Bell states---expressed in the Fock basis as $\ket{\Psi^\pm} \propto \ket{1_{A_0}1_{B_1}} \pm \ket{1_{A_1}1_{B_0}}$ in this design---produce unique coincidence patterns in the output modes and can be unambiguously identified.

Significantly, the design of such a BSA in the frequency domain introduces nuances regarding logical encoding definitions that are typically of minimal concern in the path or polarization paradigms. For example, due to the ready availability of inexpensive, passive components that can transform from any polarization basis to another, a $\ket{\Psi^\pm}$-state BSA can be adjusted to respond to a different pair of Bell states---or, equivalently, the same two Bell states under an alternative logical mapping---simply by incorporating an appropriate sequence of waveplates. For example, a type-II fusion gate that measures the $\ket{\Phi^+}$ and $\ket{\Psi^+}$ Bell states defined in the rectilinear polarization basis can be realized by replacing the nonpolarizing beamsplitter of a standard BSA with a polarizing beampslitter surrounded by half-wave plates% in each input and output port
~\cite{Browne2005}. Analogous basis transformations are attainable within frequency-bin encoding as well, yet because such transformations require additional EOMs and pulse shapers to realize, certain logical encodings may prove much more efficient to implement than others, in terms of total QFP resources (number of components and bandwidth). Accordingly, judicious placement of frequency modes can prove critical in the synthesis of multiphoton QFP gates, as evidenced by previous designs for the controlled-$Z$~\cite{Lukens2017} and controlled-NOT~\cite{LuCNOT} that attain high fidelities with appreciably fewer QFP elements than the $\mathcal{O}(N)$ worst-case scaling for an $N$-mode unitary~\cite{Lukens2017}. Irrespective of QFP gate design, relative photon arrival must be synchronized to within a fraction of the RF period while the requirements on detector jitter are considerably relaxed.

\begin{figure*}[tb!]
\centering
\includegraphics[width=5in]{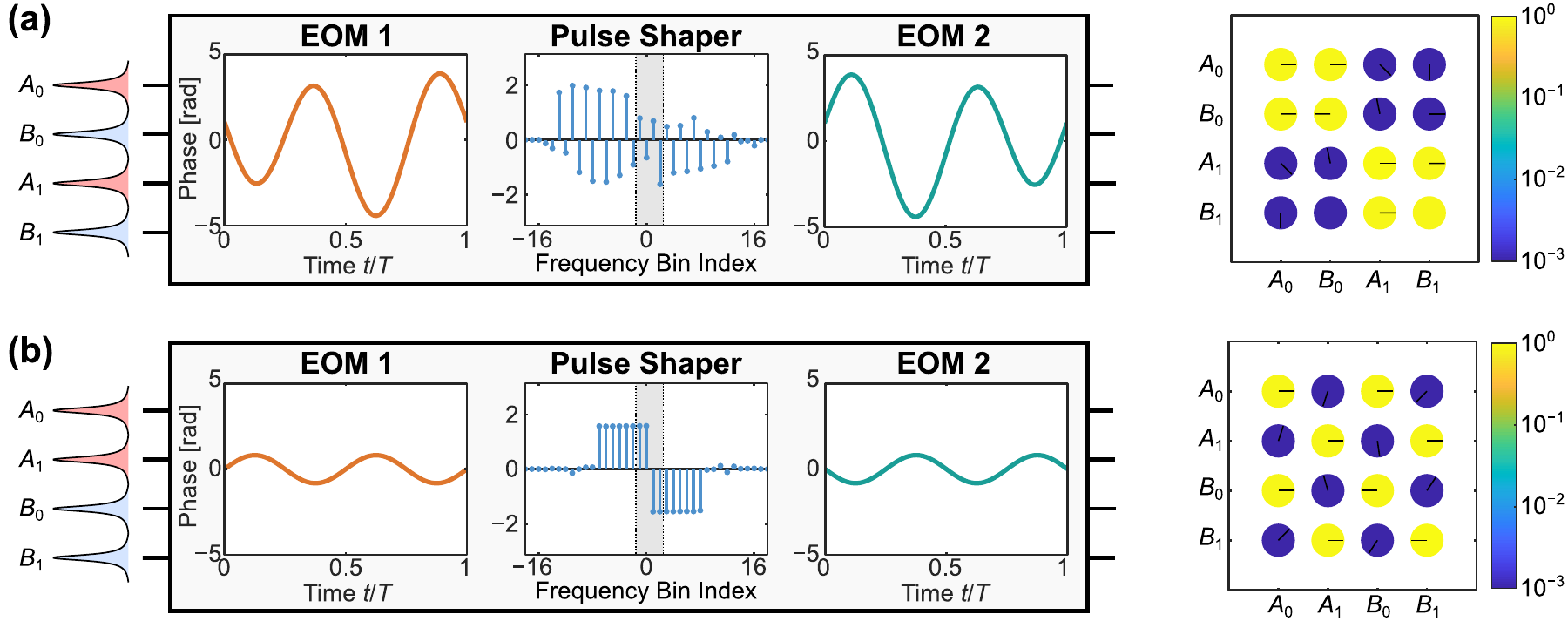} 
\caption{Optimal QFP designs for frequency-bin BSA. (a)~Interleaved qubits [encoding~(i)]. (b)~Adjacent qubits [encoding~(ii)]. The temporal and phase modulation patterns correspond to each component in the setup of Fig.~\ref{fig1}(c); the four computational modes are denoted with gray shading in the pulse shaper plot. The mode transformation matrices are depicted to the right, where the amplitude (phase) of each element is represented by the color (radial line) of the corresponding circle.}
\label{fig2}
\end{figure*}

In general, there need not exist an integer relationship between the \emph{intra}-qubit frequency separation (e.g., between modes $A_0$ and $A_1$) and the \emph{inter}-qubit spacing (e.g., between modes $A_0$ and $B_0$). Indeed, the use of two unrelated frequency grids could even prove useful in reducing crosstalk between parallel operations. However, in order to connect our experiment to situations typical for biphoton frequency combs, we assume a fixed frequency grid in our initial design. Specifically,
%As a starting point, 
consider as resources four adjacent modes centered at the frequencies $\{\omega_{-1},\omega_0,\omega_1,\omega_2\}$ where $\omega_n = \omega_0 + n\Delta\omega$. The frequency-bin transformation corresponding to a $\ket{\Psi^\pm}$-state BSA depends on whether the qubits are placed in either (i) interleaved or (ii) adjacent fashions. In encoding~(i), we can define the logical modes as $(A_0,A_1)=(\omega_{-1},\omega_1)$ and $(B_0,B_1)=(\omega_0, \omega_2)$; for encoding~(ii), the adjacent qubit definitions $(A_0,A_1)=(\omega_{-1},\omega_0)$ and $(B_0,B_1)=(\omega_1, \omega_2)$ apply. Then, the required unitaries operating on the annihilation operators associated with each bin $(\hat{a}_{-1}, \hat{a}_0, \hat{a}_1, \hat{a}_2 )^T$ are
\begin{equation}
\label{eq:unitaries}
U^{\mathrm{(i)}} = \tfrac{1}{\sqrt{2}} \begin{psmallmatrix} 1 & 1 & 0 & 0 \\ 1 & -1 & 0 & 0 \\ 0 & 0 & 1 & 1 \\  0 & 0 & 1 & -1 \end{psmallmatrix}; \;\;\;\;
U^{\mathrm{(ii)}} = \tfrac{1}{\sqrt{2}} \begin{psmallmatrix} 1 & 0 & 1 & 0 \\ 0 & 1 & 0 & 1 \\ 1 & 0 & -1 & 0 \\  0 & 1 & 0 & -1 \end{psmallmatrix}.
\end{equation}
To compare the feasibility of each encoding, we determine the optimal pulse shaper and EOM settings for the QFP in Fig.~\ref{fig1}(c) using particle swarm optimization (PSO)~\cite{Kennedy1995}. Defining $W$ as the actual $4\times 4$ mode transformation for a specific QFP configuration, PSO attempts to minimize the cost function $C=\cP_W \log_{10}(1-\cF_W)$, where the fidelity $\cF_W=|\Tr (W^\dagger U)|^2 /(16\cP_W)$ and success probability $\cP_W=\Tr (W^\dagger W)/4$ are defined in the modal sense---i.e., with respect to the desired mode unitary $U\in\{U^{\mathrm{(i)}},U^{\mathrm{(ii)}}\}$.

The optimal solutions for each encoding appear in Fig.~\ref{fig2}, which show the specific EOM modulation patterns plotted over the fundamental period $T=\frac{2\pi}{\Delta\omega}$, pulse shaper phase shifts, and complete transformation matrix $W$. We use phasor notation to represent the complex elements of the synthesized unitaries in Fig.~\ref{fig2}; the color signifies the amplitude, with the scale bar normalized to the maximum value in each matrix ($0.6831$ for $U^{\mathrm{(i)}}$ and $0.6978$ for $U^{\mathrm{(ii)}}$), and the radial line marks out the phase. The QFP is able to realize both unitaries with high fidelity ($\cF_W=1-10^{-6}$ for both) and success probability ($\cP_W=0.9310$ for $U^{\mathrm{(i)}}$ and $\cP_W=0.9739$ for $U^{\mathrm{(ii)}}$)---as evidenced by the exceptional agreement with respect to the ideal unitaries for both amplitudes and phases. Yet despite these similarities, the adjacent qubit encoding [Fig.~\ref{fig2}(b)] entails a significantly smaller maximum temporal phase deviation (0.8283~rad) compared to the interleaved encoding (4.426~rad) and consists entirely of a pure sinewave at the harmonic frequency $2\Delta\omega$, thus eliminating the need to combine two RF tones. In fact, this solution is precisely that of two frequency-bin beamsplitters following the design of Refs.~\cite{Lu2018,LuQSIM}, each operated at twice the fundamental spacing and shifted by one bin with respect to the other. On the other hand, the solution of Fig.~\ref{fig2}(a) is not clearly related to previous frequency beasmplitters, which makes sense: the fidelity of two parallel beamsplitters in Ref.~\cite{Lu2018} was found to drop rapidly with guardband separations less than two bins, so a significantly different design is required to reduce crosstalk for the contiguous beamsplitters in $U^{\mathrm{(i)}}$ here.

\begin{figure}[tb!]
\centering
\includegraphics[width=3.4in]{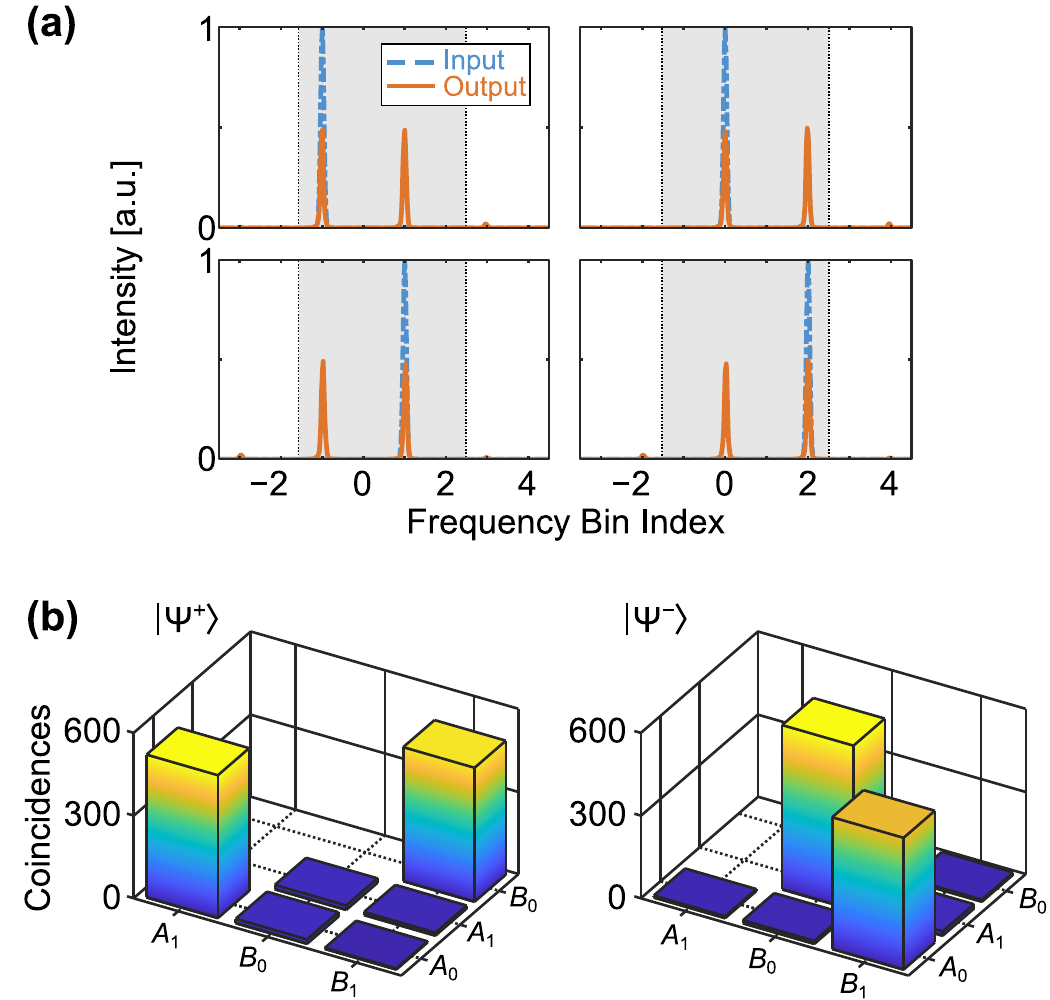} 
\caption{Experimental results for frequency-bin BSA. (a)~Measured spectra when excited by monochromatic classical inputs, obtained from an optical spectrum analyzer. Gray boxes outline the four computational modes. (b)~Coincidences obtained over all six combinations of output bins when probed by $\ket{\Psi^\pm}$ entangled states.}
\label{fig3}
\end{figure}

Given the greater simplicity of the QFP solution for encoding~(ii), we therefore focus on adjacent qubit definitions for experimental implementation. Taking a separation of $\Delta\omega/2\pi = 20$~GHz and $\omega_0/2\pi \approx 192.2$~THz, we implement the QFP solution in Fig.~\ref{fig2}(b); 
%we first characterize the QFP transformation with a classical electro-optic frequency comb~\cite{Rahimi2013, ....}, finding experimental mode fidelity $\cF_W=0.XXXX\pm0.XXXX$ and success probability $\cP_W=0.XXXX\pm0.XXXX$. 
Fig.~\ref{fig3}(a) plots measured input/output spectra for classical single-frequency inputs, highlighting independent and balanced mixing between bins $(A_0,B_0)=(\omega_{-1},\omega_1)$ and between $(A_1,B_1)=(\omega_0,\omega_2)$. For testing this frequency-bin BSA, we generate entangled photon pairs through type-0 parametric down-conversion in a periodically poled lithium niobate ridge waveguide that is pumped by a continuous-wave laser at twice the frequency of the spectral center $\omega_0 +\Delta\omega/2$. %$349~THz. %$ 2(\omega_0 + \Delta\omega/2) = 38X.XXX$ THz. 
We carve out 20~GHz-spaced frequency bins using a fiber-pigtailed etalon with an intensity full-width at half-maximum of 0.8~GHz. A subsequent pulse shaper blocks all but the four frequency bins \{$A_0$, $A_1$, $B_0$, $B_1$\} which %---when dispersion can be neglected---
ideally produces the state $|\Psi^{+}\rangle \propto \ket{1_{A_0}1_{B_1}} + \ket{1_{A_1}1_{B_0}}$; by applying a $\pi$ phase shift onto frequency bin $B_0$ as well, $|\Psi^{-}\rangle \propto \ket{1_{A_0}1_{B_1}} - \ket{1_{A_1}1_{B_0}}$ can be produced. (We did not examine the positively correlated $\ket{\Phi^{\pm}}\propto \ket{1_{A_0}1_{B_0}} \pm \ket{1_{A_1}1_{B_1}}$ Bell states, which cannot be distinguished in our setup due to the well-known 50\% optimal efficiency for vacuum-assisted linear-optical BSAs~\cite{Luetkenhaus1999, Calsamiglia2001}.) After traversing the QFP, the output is frequency-demultiplexed and routed to two superconducting nanowire single-photon detectors. Coincidence counts for all six detector combinations---$A_0A_1$, $A_0B_0$, $A_0B_1$, $A_1B_0$, $A_1B_1$, and $B_0B_1$---are collected within a 1.5~ns window and integrated for a total of 120~seconds. %We did not examine the $|\Phi^{\pm}\rangle$ Bell states experimentally, which would correspond to positive frequency correlations in our logical encoding. While this class of Bell states can be identified by observation of photon bunching, the $|\Phi^{+}\rangle$ and the $|\Phi^{-}\rangle$ states cannot be unambiguously distinguished from one another, which ultimately limits the optimal efficiency of standard, linear-optical BSAs to $50\%$~\cite{Calsamiglia2001}.

Experimental results are presented in Fig.~\ref{fig3}(b). For the $|\Psi^{+}\rangle$ input, coincidences register between the two frequencies corresponding to the original idler modes ($A_0A_1$) or the original signal modes ($B_0B_1$), as expected from theory~\cite{Mattle1996}. On the other hand, the $|\Psi^{-}\rangle$ state results in coincidences between one of the original idler modes and one of the original signal modes ($A_0B_1$ or $A_1B_0$), thereby allowing unambiguous differentiation of $|\Psi^{+}\rangle$ and $|\Psi^{-}\rangle$. We calculate the discrimination accuracy $N_C/(N_C + N_I)$, where $N_C$ and $N_I$ correspond, respectively, to the sum of the two correct measurement results and the sum of the two incorrect results---i.e., those which misidentify the state---when the input is $|\Psi^{+}\rangle$ or $|\Psi^{-}\rangle$. We compute accuracies of $(98.1\pm0.04)\%$ and $(98.6\pm0.04)\%$ for $|\Psi^{+}\rangle$ and $|\Psi^{-}\rangle$, respectively, assuming Poissonian error bars and without any subtraction of accidentals.  
%To estimate the accuracy of our BSA while accounting for the finite statistics of our experiment, we consider a simple Bayesian model that assigns probabilities $\{p_{A_0A_1}, p_{A_0B_1}, p_{A_1B_0}, p_{B_0B_1}\}$ to the four-outcome subspace of interest for state discrimination; then the accuracies can be defined as $a_+ = p_{A_0A_1}+ p_{B_0B_1}$ and $a_- = p_{A_0B_1} + p_{A_1B_0}$ for the $\ket{\Psi^+}$ and $\ket{\Psi^-}$ cases, respectively. Taking a uniform Dirichlet distribution as prior for the four probabilities in each case and modeling the counts as a multinomial distribution,  straightforward inference~\cite{} returns $a_+ = a_- = 0.98\pm 0.02$, where error bars correspond to the standard deviation of the Bayesian posterior. Such high discrimination accuracies---computed without accidental subtraction---offer additional quantitative confirmation of the exceptional performance of our frequency-bin BSA. % beyond the visual indication in Fig.~\ref{fig3}(b).

The concept of frequency mixing-based BSAs is not limited to frequency-encoded quantum information and can be extended to Bell state measurements of spectrally distinguishable photons encoded in other degrees of freedom. For example, in the case of spectrally distinguishable polarization qubits, one would implement two frequency beamsplitters within a polarization-diversity scheme, and for time-bin qubits a single frequency beamsplitter would suffice.

Moreover, the QFP approach is not the only way to synthesize BSAs based on frequency mixing. A single EOM is sufficient to realize a probabilistic frequency-bin beamsplitter, whereby photonic energy scattered into adjacent sidebands is lost and not compensated through subsequent stages; this simpler design has been employed in single-photon entanglement swapping protocols~\cite{Lan2007}, as well as frequency-bin Hong--Ou--Mandel interference experiments~\cite{Imany2018c, Kashi2021}, suggesting promise in a complete BSA. A coupled-cavity--based frequency-bin beamsplitter~\cite{Zhang2019, Hu2020} eliminates the additional sidebands produced by a nonresonant EOM and therefore provides a highly compact, integrated platform for future BSAs. Finally, through appropriate design of classical pump fields and phase-matching conditions, previously demonstrated frequency beamsplitters based on $\chi^{(2)}$~\cite{Kobayashi2016} and $\chi^{(3)}$~\cite{Clemmen2016, Joshi2020b} optical nonlinearities provide opportunities for frequency-bin BSAs bridging large (THz and beyond) spectral separations.
%One can also realize these operations using integrated optical components like coupled-cavity electro-optic modulators~\cite{Hu2020} or techniques like Bragg scattering four-wave mixing~\cite{Joshi2020b} to bridge large frequency differences. %In addition, probabilistic frequency interference with a single modulator has been used in implementations of single-photon entanglement swapping protocols~\cite{Lan2007}. 
Irrespective of the particular physical implementation, BSAs based on frequency mixing have the potential to support heterogeneous nodes and dense spectral multiplexing on quantum networks without imposing additional limits on entanglement fidelity or the entanglement generation rate.

\medskip

\noindent\textbf{Funding.} U.S. Department of Energy, Office of Science, Office of Advanced Scientific Computing Research (Early Career Research Program); National Science Foundation (NSF) (2034019-ECCS, 1747426-DMR); AFRL Prime Order No. FA8750-20-P-1705

\medskip

\noindent\textbf{Acknowledgments.} Some preliminary results for this article were presented at IPC 2020 as paper number PD5 and CLEO 2021 as paper number FTu1N.5. We thank AdvR for loaning the PPLN ridge waveguide. A portion of this work was performed at Oak Ridge National Laboratory, operated by UT-Battelle for the U.S. Department of Energyy under contract no. DE-AC05-00OR22725

\medskip

\noindent\textbf{Disclosures.} The authors declare no conflicts of interest.

\end{document}